\documentclass[mathleft
]{an}
\usepackage{graphicx}
\usepackage{times}
\begin{document}

\Pagespan{789}{}
\Yearpublication{2006}%
\Yearsubmission{2005}%
\Month{11}%
\Volume{999}%
\Issue{88}%

\title{Testing of the dwarf galaxy content and the evolutionary status of nearby groups of galaxies}

\author{J. Vennik\inst{1}\fnmsep\thanks{\email{vennik@aai.ee}\newline}
\and  U. Hopp\inst{2,3}
}
\titlerunning{Nearby groups of galaxies}
\authorrunning{J. Vennik \& U. Hopp}
\institute{
Tartu Observatory, 61602 T\~oravere, Tartumaa, Estonia
\and 
Universit\"ats--Sternwarte M\"unchen, Scheiner Str. 1, D 81679 M\"unchen, Germany
\and 
Max--Planck--Institut f\"ur Extraterrestrische Physik,D 85741 Garching, Germany}

\received{}
\accepted{}


\abstract{
  We carried out visual and parametric searches for dwarf galaxies in
  five loose groups of galaxies. Follow-up spectroscopy with the HET
  has shown a 50\% success rate of morphological
  selection. The evolutionary status of the studied groups differs:
  while the NGC 6962 group has a partially relaxed core, surrounded by an
  infall region, the NGC 5005/5033 group and the IC 65 group, which consist only of 
  late-type galaxies, are probably still assembling.
}

\maketitle

\section{Introduction}

Groups of galaxies may represent sites for a preprocessing stage of
cluster galaxies through some varieties of gravitational (tidal) and
hydrodynamical mechanisms acting in groups.  Dwarf galaxies are
expected to reflect the environmental influence most prominently.  We
have selected dwarf galaxy candidates in a sample of nearby loose
groups of galaxies, basing on morphological criteria including surface
photometry.
The main aim of our project is to investigate the impact of the group
environment on the evolution of the dwarfs as well as on the evolution
of the major group members due to dwarf galaxy accretion. 
Here we report the preliminary results
obtained for three groups around the principal galaxies IC 65, NGC
6962, and NGC 5005/5033, respectively, using the archival data from
the SDSS, complemented with our own imaging studies at Calar Alto,
and spectroscopy with the Hobby-Eberly Telescope (HET).

\section{Candidate selection and membership confirmation progress report}

\subsection{Candidate selection on the DPOSS and with our own imaging data (IC 65 group)}

We carried out a parametric search for new dwarf galaxies in the area
of the IC 65 group with SExtractor software on the DPOSS frames. For
the preliminary classification of detected galaxies we used (a) the
Binggeli's (1994) empirical relation between the central surface
brightness (SB) and absolute magnitude, and (b) the empirical light
concentration parameter (Trentham et al. 2001) , versus SExtracted
colour index. As result, we have selected four LSB irregular galaxies
with lowest light concentration and the bluest colour (0.45 $< B-R<$
1.05). We have obtained CCD $B, R$ and $I$ imaging data with the Calar
Alto 1.23m telescope and performed detailed surface photometry of all
secure and possible members of the IC 65 group (Vennik \& Hopp 2008).
The final classification of the new candidate galaxies has been
carried out according to their SB (LSB, HSB), morphology (irregular,
regular) and colour (blue, red).

\subsection{Candidate selection with the SDSS data (NGC 6962 and NGC 5005/5033 groups)}

The area of these two groups is covered by the Sloan Survey and we
have searched for new dwarf member candidates of these groups using
the homogeneous imaging data of the SDSS. We have extracted the list
of the dwarf galaxy candidates from the $PhotoObjAll$ catalog using
the selection criteria as follows: the effective SB $\mu_{\mathrm {eff}} >
22.0$ g~arcsec$^{-2}$; light concentration $C = petroR50/petroR90 >
0.4$; size $isoA > 15$ arcsec. All pre-selected dwarf candidates have
been visually inspected on the SDSS frames and final classification
has been made on their morphological and colour grounds. 

For NGC 6962 group of galaxies, we retrieved 29 probable (and 44
possible) members of the group within 1 Mpc around the parent galaxy
NGC 6962. The comparison of angular 2D correlation properties of true
(confirmed by redshift) and candidate members of the NGC 6962 group
shows that the ensemble of probable members may consist a mix of $\sim
55 \%$ of true members and $\sim 45 \%$ of randomly distributed
"field" objects (Vennik \& Tago 2007).

For the NGC~5005/5033 group of galaxies, our survey resulted into
5 probable (and 8 possible) dwarf galaxies wit\-hin 1.5 Mpc centered on the
two major group members. 

\subsection{HET spectroscopy -- candidate confirmation progress report}

\begin{figure}[h]
\includegraphics[width=50mm,height=75mm,angle=-90]{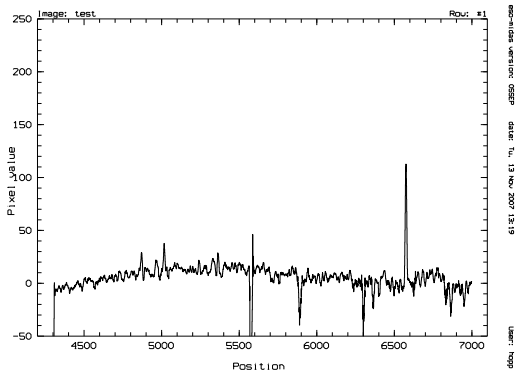}
\caption{
  HET LRIS long slit spectrum of a new dwarf member of the IC 65 group
  showing the typical emission lines of HII galaxies (and night sky
  residuals).}
\label{fig1}
\end{figure}

\begin{figure}[h]
\vspace{-15mm} 
\hspace{-3mm}
\includegraphics[width=30mm,height=52mm]{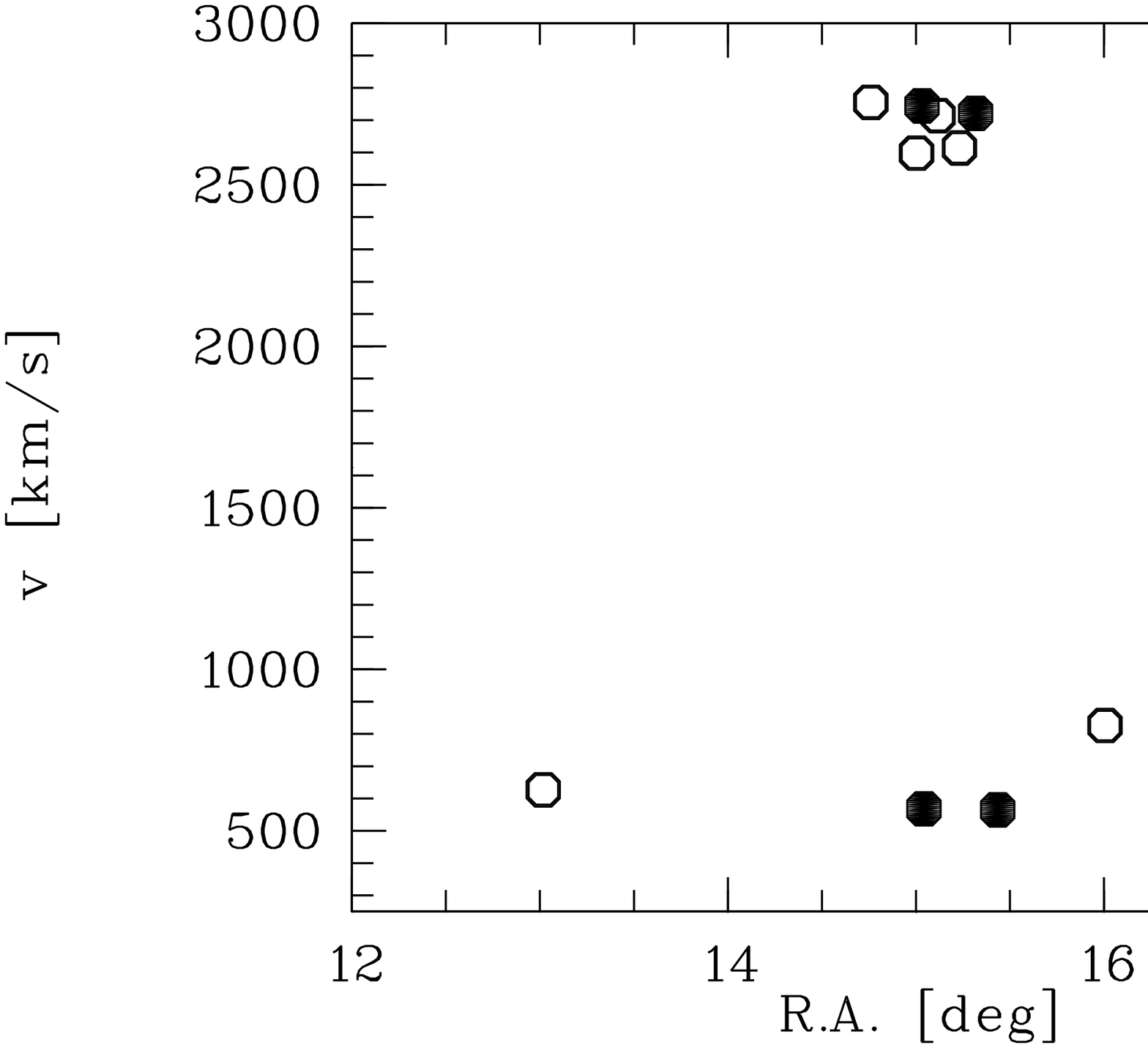}
\hspace{11mm}
\includegraphics[width=30mm,height=52mm]{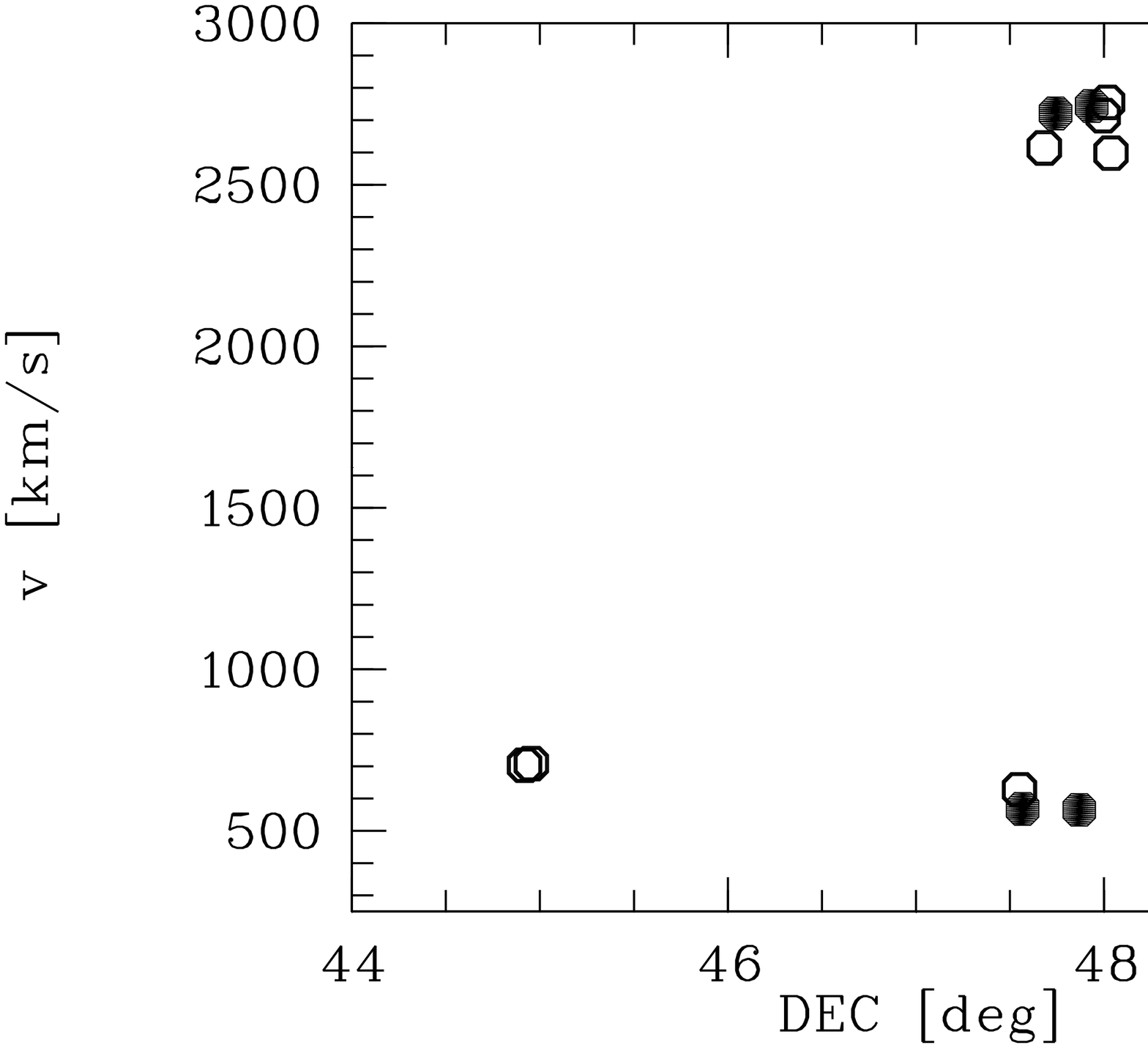}
\caption{Redshift distribution towards the IC~65 group: 
filled symbols - new redshifts measured with the HET; 
open symbols - redshifts obtained from the NED.
The IC 65 group with two new dwarf members, 
confirmed by HET, has a mean heliocentric redshift 
2670 km/s. The other two newly detected dwarf galaxies with  
$v \sim$ 600 km/s are probably members of the NGC 278 
group of galaxies.  
}
\label{fig2}
\end{figure}

Follow-up spectroscopy of the selected dwarf galaxy \\
candidates has
been carried out with the HET low resolution spectrograph
(2$^{\prime\prime} \times 4^\prime$ long slit, Grism with 5\AA~ or
2\AA~ per pixel, in the range of 430 to 800 nm), with aim to obtain
redshifts and SED classification (Fig. \ref{fig1}). During several observing
runs in 2007 and 2008 19 redshifts were obtained towards five groups
with following results: 8 confirmed new members, 2 confirmed
background objects, 9 member candidates rejected (either background or
foreground).  Among the four member candidates of the IC 65 group two
are true dwarf members of this group and the other two are dwarf
galaxies but members of the newly identified foreground group around
the NGC 278 (Fig. \ref{fig2}). All four new member candidates of the NGC 6962
group, observed with the HET, are confirmed as true members of this group.

Thus, in total, the morphological classification has a success rate of
52\% according to the spectroscopic follow-up. Almost all of the confirmed
dwarf members are of late morphological type and show emission lines
pointing to ongoing star formation activity.

\section{The NGC 6962 group}

\begin{figure}
\vspace{3mm}
\includegraphics[width=70mm,height=50mm]{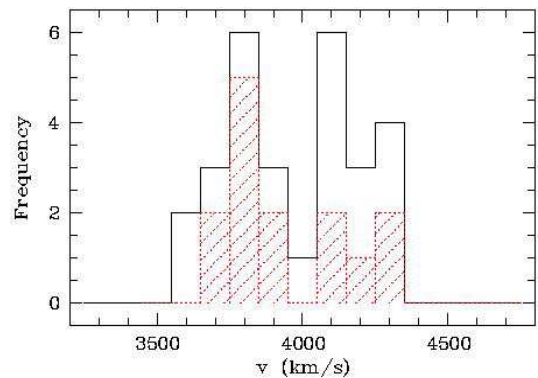}
\caption{The NGC 6962 group velocity distribution is bi-modal with peaks at 3765$\pm$25 km/s and
4171$\pm$27 km/s. Both velocity-subgroups almost overlap in the sky and contain nearly equal fraction
of early-type (shaded histogram) and late-type galaxies.}
\label{fig3}
\end{figure}

\begin{figure}[h]
\vspace{-30mm}
\includegraphics[width=60mm,height=90mm]{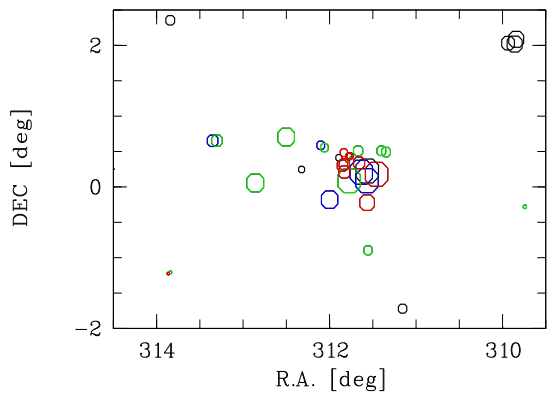}
\vspace{5mm} 
\caption{Distribution of the NGC 6962 group members within $\sim$ 3 Mpc around the parent galaxy 
(and are colour-coded blue-green-red according to their decreasing current star-forming
activity, classified by us as in Tully \& Trentham (2008)). The
bubble sizes are scaled according to the Dressler-Shectman (1988) DS-test for substructure -- isolated
congregation of large bubbles is indicative of a kinematically distinct subgroup.}
\label{fig4}
\end{figure}

The NGC 6962 group is a dense group at the distance of 56 Mpc. It is well isolated, both in the sky and in velocity-space. There are 38 galaxies with 
concordant redshifts wit\-hin $r_0$ = 2.6 Mpc (zero-velocity radius) around the parent galaxy, listed 
in the SDSS and/or in the NED. By now, four new member candidates have been confirmed by the HET spectroscopy. 
The group has well-defined core - halo structure, with the core (or second turn-around) radius 
$r_{\mathrm 2t} \simeq$ 0.73 Mpc with 26 secure members, and halo (or infall region) extending to 
$r_0$ = 2.6 Mpc. The group kinematics is characterized by the bi-modal velocity histogram 
(Fig.~ \ref{fig3}), and by several congregations of big bubbles in the Dressler-Shectman (1988) DS-test, 
which are indicative of kinematically distinct subgroups (Fig.~ \ref{fig4}). 
The NGC 6962 group shows a classical morphology-density (and spectral-densi\-ty) correlation with 
early/passive types (E, E+ pec, S0) mos\-tly in the core, and late/star-forming types in the halo.\\ 
The group luminosity function is essentially flat in the range of $-21 \leq M_g \leq -16$. \\
{\it Evolutionary status of the NGC 6962 group:} evidence of segregation, declining velocity dispersion 
profile, and relatively short crossing time ($t_{\mathrm {c}}H_0 \simeq 0.14$) of the core may be indicative 
of a partially relaxed group core. 
According to the ROSAT Survey it is an X-ray dark group, i.e. the intra-group gas, if 
stripped in early stages of the group evolution from spiral galaxies by mutual interactions, 
is still in a relatively cold stage. The disturbed E+ pec galaxies may be late merger remnants.

\section{The NGC 5005/5033 group}

\begin{figure}
\vspace{-1cm}
\includegraphics[width=60mm,height=80mm]{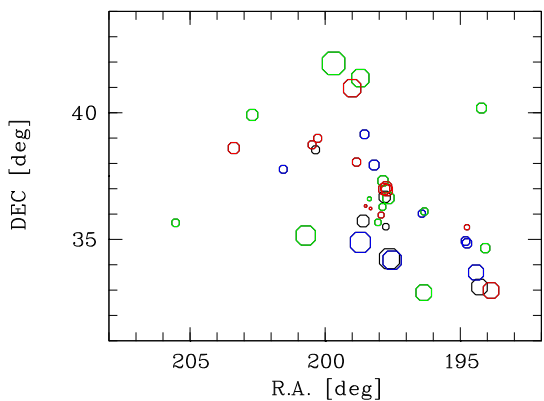}
\caption{The distribution of the NGC 5005/5033 group members within $\sim$ 1.5 Mpc around the group 
barycenter. The bubble sizes are scaled according to the DS-test (as in the Fig. 4). }
\label{fig5}
\end{figure}

This is a moderately isolated loose aggregation of 41 galaxies (SDSS
\& NED) within 6 degrees ($\sim$1.5 Mpc) around the group barycenter,
at the distance of 13.7 Mpc. Two new dwarf members were confirmed by
our HET spectroscopy.  The group consists of only late-type galaxies
(later than Sab), and is dominated by a wide pair of luminous spirals,
namely NGC 5005 ($M_B = -21.4$) and NGC 5033 ($M_B = -20.8$). The group
includes several kinematically distinct subgroups of faint, actively
star forming galaxies (e.g. around UGC 8246 and UGC 8365), possibly
falling to the group along the filament(s) (compare Fig.~\ref{fig5}).
The group velocity dispersion profile is rising with increasing radius. \\
{\it Evolutionary status:} the preliminary analysis let us suppose that this group is 
probably still assembling and star-forma\-tion in gas-rich late-type galaxies may be driven by mutual 
interactions, particularly in kinematically distinct subgroups. 

\section{Summary}

For a sample of nearby loose groups, we complemented the member ship lists
with new dwarf galaxies where the group membership has been confirmed 
by HET spectroscopy. 
The preliminary analysis of the NGC 6962 and NGC 5005/5033 groups show, 
that these two groups differ in their spatial structure, kinematics and 
morphological/spectral content, \\ which possibly indicate their different 
evolutionary status. The IC 65 group has been analyzed in more detail 
previously (Vennik \& Hopp 2007, 2008), and the evolutionary status of this group can be 
summarized being at a relatively early stage of its collapse.

\acknowledgements

The research of JV has been supported by the Estonian Science
Foundation grants 6106 and 7765.  We are grateful to the organizers of
JENAM2008.  This study has made use of the NASA/IPAC Extragalactic
Database (NED), the STScI Digitized Second Palomar Obsrvatory Sky Survey (DPOSS), and the Sloan
Digital Sky Survey (SDSS). The Hobby-Eberly Telescope (HET) is a joint
project of the University of Texas at Austin, the Pennsylvania State
University, Stanford University, Ludwig-Maximilians-Universit\"at
M\"unchen, and Georg-August-Universit\"at
G\"ottingen. The HET is named in honor of its principal
benefactors, William P. Hobby and Robert E. Eberly.


\end{document}